\documentclass[12pt,showpacs,showkeys,amsmath,amssymb]{revtex4}
\usepackage{amsmath,amsfonts,amsthm,amscd,amssymb,latexsym}
\usepackage{bm}
\usepackage{dcolumn}
\usepackage{graphicx}
\usepackage{epstopdf}
\usepackage{color}
\usepackage{epsf}
\usepackage{epsfig}
\usepackage{graphicx, epic, eepic, color}

\newcommand{\beq}{\begin{equation}}
\newcommand{\eeq}{\end{equation}}

\def\@{\partial_}
\def\be{\begin{equation}}
\def\ee{\end{equation}}

\def\negenspace{\kern-1.1em}

\def\sqr#1#2{{\vcenter{\hrule height.#2pt\hbox{\vrule width.#2pt
height#1pt \kern#1pt \vrule width.#2pt}\hrule height.#2pt}}}

\begin{document}

\title{ Anisotropic Cosmology in the Local Limit of Nonlocal Gravity}

\author{Javad \surname{Tabatabaei}$^{1}$}
\email{smj_tabatabaei@physics.sharif.edu}
\author{Abdolali \surname{Banihashemi}$^{1}$}
\email{abdolali.banihashemi@sharif.edu}
\author{Shant \surname{Baghram}$^{1}$}
\email{baghram@sharif.edu}
\author{Bahram \surname{Mashhoon}$^{2,3}$} 
\email{mashhoonb@missouri.edu}

\affiliation{
$^1$Department of Physics, Sharif University of Technology, Tehran 11155-9161, Iran\\
$^2$School of Astronomy,
Institute for Research in Fundamental Sciences (IPM),
Tehran 19395-5531, Iran\\
$^3$Department of Physics and Astronomy,
University of Missouri, Columbia,
Missouri 65211, USA \\
}

\date{\today}

\begin{abstract}
Within the framework of the local limit of nonlocal gravity (NLG), we investigate a class of Bianchi type I  spatially homogeneous but anisotropic cosmological models. The modified field equations are presented in this case and some special solutions are discussed in detail. This modified gravity theory contains a susceptibility function $S(x)$ such that general relativity (GR) is recovered for $S = 0$. In the modified anisotropic cosmological models, we explore the contribution of $S(t)$ and its temporal  derivative to the local anisotropic cosmic acceleration. The implications of our results for observational cosmology are briefly discussed. 
\end{abstract}


\maketitle


\section{Introduction}

In the current $\Lambda$CDM model of cosmology, the energy content of the universe consists of about  70\% dark energy, about 25\% dark matter, and about 5\% visible matter. The model is based on the standard spatially homogeneous and isotropic Friedmann--Lema\^{\i}tre--Robertson--Walker (FLRW)  cosmological solutions of Einstein's general relativity theory~\cite{Einstein}. The \emph{dark} features, whose nature and origin are unknown, provide the motivation to modify and extend general relativity (GR) on galactic scales and beyond in order to account for observational data purely on the basis of the gravitational physics of the extended GR without recourse to dark ingredients. 

To modify the current benchmark model of cosmology, we consider nonlocal gravity (NLG) theory~\cite{Hehl:2008eu,Hehl:2009es, BMB},  a classical history-dependent generalization of GR that bears a formal resemblance to the nonlocal electrodynamics of media~\cite{Hop, Poi, Jackson, L+L, HeOb, VanDenHoogen:2017nyy}. It is important to digress here and mention that there are indeed various other approaches to nonlocal gravitation. For the sake of brevity, we only refer to some examples and their cosmological implications. Nonlocally modified extensions of GR can be generated by the addition of functions of $\Box$, as in infinite derivative theories, or functions of $\Box^{-1}$ to the Einstein-Hilbert action. Here, $\Box$ is the d'Alembert-Beltrami operator. Cosmological implications of such theories have been investigated by a number of authors in connection with dynamic dark energy and accelerated expansion of the universe; for instance, see~\cite{Maggiore:2014sia, Capozziello:2021krv} and the references cited therein. Moreover, cosmological solutions of nonlocal infinite derivative theories have been studied that involve anisotropic bouncing models~\cite{Kumar:2021mgc} or an interplay between dark matter and dark energy~\cite{Dimitrijevic:2019pct}. Quantum field theory provides the motivation for a different class of nonlocal theories of gravitation. Higher curvature nonlocal gravity theories have recently been reviewed in~\cite{Koshelev:2023elc} and a generalized nonlocal quantum gravity theory has been formulated within the framework of inflationary cosmology. On the other hand, the phenomenological approach of Deser and Woodard has been based on an effective quantum gravitational action and has been designed to explain cosmic acceleration without dark energy; see~\cite{Deser:2019lmm} and the references cited therein. Furthermore, primordial bouncing cosmology and anisotropy have been investigated within the framework of the Deser-Woodard nonlocal gravity model in~\cite{Chen:2019wlu}. 

We now return to our classical model of nonlocal gravity that is patterned after the nonlocal electrodynamics of media.  Nonlocal gravity (NLG) is a tetrad theory, where the gravitational potentials are given by the 16 components of a preferred orthonormal tetrad frame field. The extended geometric framework of NLG is based on the Weitzenb\"ock connection~\cite{We}, which renders the spacetime a parallelizable manifold. Within the framework of teleparallelism~\cite{Maluf:2013gaa, Aldrovandi:2013wha, BlHe, Itin:2018dru}, it is possible to express GR using the Weitzenb\"ock torsion tensor. This teleparallel equivalent of general relativity (TEGR) is a gauge theory of the Abelian group of spacetime translations~\cite{Cho}. The formal similarity between TEGR and electrodynamics can be employed to introduce nonlocality into GR via constitutive kernels~\cite{Hehl:2008eu, Hehl:2009es}.  In NLG, the gravitational field is local, but the theory involves an average of the field over past events resulting in 16 partial integro-differential field equations~\cite{Puetzfeld:2019wwo, Mashhoon:2022ynk}. No exact nontrivial solution of NLG is known at present~\cite{Bini:2016phe}; however, the linear regime of the theory has been extensively studied. Nonlocal gravity, in its Newtonian regime, simulates dark matter. It is therefore possible to account for the gravitational effects in the solar system as well as in nearby galaxies and clusters of galaxies~\cite{Rahvar:2014yta, Chicone:2015coa, Roshan:2021ljs, Roshan:2022zov, Roshan:2022ypk}. A comprehensive account of these studies is contained in~\cite{BMB}.

NLG is rather intricate and to study its cosmological implications, we resort to its local limit, which is easier to analyze. In Section II, we present a brief account of the modified GR field equations in the local limit of NLG. For a more detailed treatment of this limiting situation, see~\cite{Tabatabaei:2022tbq, Tabatabaei:2023qxw}, where spatially homogeneous and isotropic (FLRW) cosmological models were investigated in this modified TEGR scheme. To explore anisotropy in the Hubble flow, we present in Section III modified gravitational field equations for a Bianchi type I class of time-dependent spatially homogeneous but  anisotropic spacetimes within the framework of the local limit of NLG. The field equations contain a susceptibility function $S(t)$ with $1+S > 0$ and $dS/dt \ne 0$ that is characteristic of the dynamic spacetime background. For $S = 0$, we recover the GR field equations.  We show that de Sitter and Kasner spacetimes are not solutions of the modified field equations unless $S(t)$ is independent of time, which is not physically reasonable. Explicit solutions of the modified field equations are studied in the next two sections.  A well-known class of dynamic solutions of GR for dust with vanishing cosmological constant is extended to the local limit of NLG in Section IV. The new solutions contain the time-dependent function $S(t)$ and allow the possibility of exploring the dependence of anisotropic acceleration on $S(t)$. These modified cosmological models are locally anisotropic but tend to the isotropic modified Einstein-de Sitter model at late times ($t \to \infty$). Similarly, 
 we study the solution of the modified field equations for a spacetime dominated by dark energy in Section V and explore  anisotropic cosmic acceleration in this cosmological model, which eventually becomes isotropic as well. The presence of $S(t)$ could be responsible for certain new ``dark" features of accelerating bulk flows in the local universe. 
 
 Anisotropy of the Hubble flow would indicate a significant departure from the presumed large-scale spatial homogeneity and isotropy of the standard FLRW cosmology. On the other hand, there is recent observational evidence in support of \emph{local} anisotropic cosmic acceleration~\cite{Colin:2019opb, Secrest:2020has, Secrest:2022uvx, Solanki:2023yoa, Perivolaropoulos:2023tdt}.   
The purpose of the present paper is to study theoretically the possible contribution of the susceptibility function $S(t)$ to anisotropic features of the Hubble flow.


\section{Local Limit of NLG}

We consider a spacetime manifold as in general relativity (GR). In an admissible system of coordinates $x^\mu$, the spacetime metric can be written as
\begin{equation}\label{1}
ds^2 = g_{\mu \nu}(x)\,dx^\mu \, dx^\nu\,.
\end{equation}
Here, Greek indices run from 0 to 3, Latin indices run from 1 to 3, and the signature of the metric is +2; moreover, we employ units such that $c = 1$.  As in GR, the world lines of free test particles and null rays are geodesics of the spacetime manifold.  We assume the existence of a preferred set of observers in this gravitational field. The observers have adapted orthonormal tetrads $e^\mu{}_{\hat {\alpha}}(x)$,
\begin{equation}\label{2}
g_{\mu \nu}(x) \, e^\mu{}_{\hat {\alpha}}(x)\, e^\nu{}_{\hat {\beta}}(x)= \eta_{\hat {\alpha} \hat {\beta}}\,,
\end{equation}
where $\eta_{\alpha \beta} = {\rm diag}(-1,1,1,1)$ is the Minkowski metric tensor. In our convention,  indices without hats are normal spacetime indices, while hatted indices indicate the tetrad axes in the local tangent space.

We employ the tetrad frame field to define the curvature-free Weitzenb\"ock connection, 
\begin{equation}\label{3}
\Gamma^\mu_{\alpha \beta}=e^\mu{}_{\hat{\rho}}~\partial_\alpha\,e_\beta{}^{\hat{\rho}}\,.
\end{equation}
Let $\nabla$ denote covariant differentiation with respect to the Weitzenb\"ock connection; then,  $\nabla_\nu\,e_\mu{}^{\hat{\alpha}}=0$, thereby the preferred tetrad frames are parallel throughout the gravitational field and provide a natural scaffolding for the spacetime manifold. The spacetime is thus a parallelizable manifold by the Weitzenb\"ock connection. In this framework of teleparallelism, two distant vectors are considered parallel if they have the same local components relative to their preferred tetrad frames. Moreover, it follows from the tetrad orthonormality relation that the Weitzenb\"ock connection is metric compatible, namely, $\nabla_\mu\,g_{\alpha \beta} = 0$. 

The difference between two connections on the same manifold is a tensor. We define the  \emph{torsion} tensor that corresponds to the Weitzenb\"ock connection by
\begin{equation}\label{4}
C_{\mu \nu}{}^{\alpha}=\Gamma^{\alpha}_{\mu \nu}-\Gamma^{\alpha}_{\nu \mu}=e^\alpha{}_{\hat{\beta}}\Big(\partial_{\mu}e_{\nu}{}^{\hat{\beta}}-\partial_{\nu}e_{\mu}{}^{\hat{\beta}}\Big)\,.
\end{equation}
In the extended GR framework, we have the Weitzenb\"ock connection as well as the symmetric Levi-Civita connection,
\begin{equation}\label{5}
^0\Gamma^\mu_{\alpha \beta} = \frac{1}{2} g^{\mu \nu}\,(g_{\nu \alpha, \beta} + g_{\nu \beta, \alpha} - g_{\alpha \beta, \nu})\,.
\end{equation}
We use a left superscript ``0" to refer to geometric quantities directly derived from the Levi-Civita connection.  The \emph{contorsion} tensor is then defined by
\begin{equation}\label{6}
K_{\mu \nu}{}^{\alpha} =\, ^{0}\Gamma^{\alpha}_{\mu \nu} - \Gamma^{\alpha}_{\mu \nu}\,,
\end{equation}
which is related to the torsion tensor through the metric compatibility of the Weitzenb\"ock connection. In fact, 
\begin{equation}\label{7}
K_{\mu \nu \rho} = \frac{1}{2}\, (C_{\mu \rho \nu}+C_{\nu \rho \mu}-C_{\mu \nu \rho})\,.
\end{equation}

The Levi-Civita connection given by the Christoffel symbol is the sum of the Weitzenb\"ock connection and the contorsion tensor.  One can therefore express the Einstein tensor $^0G_{\mu \nu}$  and the gravitational field equations of GR in terms of the teleparallelism framework resulting in the teleparallel equivalent of GR, namely, TEGR~\cite{BMB}. Indeed, we find
\begin{align}\label{8}
 {^0}G_{\mu \nu}=\frac{\kappa}{\sqrt{-g}}\Big[e_\mu{}^{\hat{\gamma}}\,g_{\nu \alpha}\, \frac{\partial}{\partial x^\beta}\,\mathfrak{H}^{\alpha \beta}{}_{\hat{\gamma}}
-\Big(C_{\mu}{}^{\rho \sigma}\,\mathfrak{H}_{\nu \rho \sigma}
-\frac{1}{4}\,g_{\mu \nu}\,C^{\alpha \beta \gamma}\,\mathfrak{H}_{\alpha \beta \gamma}\Big) \Big]\,
\end{align}
and
Einstein's field equations expressed in terms of torsion thus become the TEGR field equations
\begin{equation}\label{9}
\frac{\partial}{\partial x^\nu}\,\mathfrak{H}^{\mu \nu}{}_{\hat{\alpha}}+\frac{\sqrt{-g}}{\kappa}\,\Lambda\,e^\mu{}_{\hat{\alpha}} =\sqrt{-g}\,(T_{\hat{\alpha}}{}^\mu + \mathbb{T}_{\hat{\alpha}}{}^\mu)\,,
\end{equation}
where $\Lambda$ is the cosmological constant and $\kappa := 8 \pi G$. Here, we define the auxiliary torsion field $\mathfrak{H}_{\mu \nu \rho}$ by means of the auxiliary torsion tensor $\mathfrak{C}_{\alpha \beta \gamma}$, namely,
\begin{equation}\label{10}
\mathfrak{H}_{\mu \nu \rho}:= \frac{\sqrt{-g}}{\kappa}\,\mathfrak{C}_{\mu \nu \rho}\,, \qquad \mathfrak{C}_{\alpha \beta \gamma} :=C_\alpha\, g_{\beta \gamma} - C_\beta \,g_{\alpha \gamma}+K_{\gamma \alpha \beta}\,.
\end{equation}
Moreover,  $C_\mu :=C^{\alpha}{}_{\mu \alpha} = - C_{\mu}{}^{\alpha}{}_{\alpha}$ is the torsion vector. As in GR, $T_{\mu \nu}$  is the symmetric energy-momentum tensor of matter. We interpret $\mathbb{T}_{\mu \nu}$ in Equation~\eqref{9} to be the traceless energy-momentum tensor of the gravitational field given by
\begin{equation}\label{11}
\mathbb{T}_{\mu \nu} := (\sqrt{-g})^{-1}\, (C_{\mu \rho \sigma}\, \mathfrak{H}_{\nu}{}^{\rho \sigma}-\tfrac{1}{4}  g_{\mu \nu}\,C_{\rho \sigma \delta}\,\mathfrak{H}^{\rho \sigma \delta})\,.
\end{equation}
This version of GR, namely, TEGR, is the gauge theory of the 4-parameter Abelian group of spacetime translations~\cite{Cho}; therefore, though nonlinear, it bears a certain resemblance to Maxwell's electrodynamics. 

In analogy with the electrodynamics of media, we can consider the torsion tensor in the form $C_{\mu \nu}{}^{\hat \alpha} = \partial_{\mu}e_{\nu}{}^{\hat{\alpha}}-\partial_{\nu}e_{\mu}{}^{\hat{\alpha}}$ to be similar to the Faraday tensor, while the relationship between $\mathfrak{H}_{\mu \nu \rho}$ and the torsion tensor in Equation~\eqref{10} can be viewed as the local constitutive relation of TEGR. Let us recall that in Maxwell's electrodynamics, the constitutive relation may change, but the field equations remain the same. We adopt the same approach for the purpose of modifying Einstein's theory.  That is, we modify TEGR by introducing a tensor $N_{\mu \nu \rho} = - N_{\nu \mu \rho}$ that changes  the constitutive relation of TEGR as follows
\begin{equation}\label{12}
\mathcal{H}_{\mu \nu \rho} = \frac{\sqrt{-g}}{\kappa}(\mathfrak{C}_{\mu \nu \rho}+ N_{\mu \nu \rho})\,.
\end{equation}
To obtain the field equations of modified TEGR,  we simply replace $\mathfrak{H}$ in Equations~\eqref{9} and~\eqref{11} by $\mathcal{H}$. The gravitational field equations of extended GR based on the new tensor field  $N_{\mu \nu \rho}$ now take the form
\begin{equation}\label{13}
 \frac{\partial}{\partial x^\nu}\,\mathcal{H}^{\mu \nu}{}_{\hat{\alpha}}+\frac{\sqrt{-g}}{\kappa}\,\Lambda\,e^\mu{}_{\hat{\alpha}} =\sqrt{-g}\,(T_{\hat{\alpha}}{}^\mu + \mathcal{T}_{\hat{\alpha}}{}^\mu)\,,
\end{equation}
where $\mathcal{T}_{\mu \nu}$ is the traceless energy-momentum tensor of the gravitational field. The gravitational energy-momentum tensor is modified by the presence of $N_{\mu \nu \rho}$; hence, we introduce a  traceless tensor $Q_{\mu \nu}$ that indicates this difference, namely, 
\begin{equation}\label{14}
\kappa\,\mathcal{T}_{\mu \nu} = \kappa\,\mathbb{T}_{\mu \nu} + Q_{\mu \nu}\,,
\end{equation}
where 
\begin{equation}\label{15}
Q_{\mu \nu} := C_{\mu \rho \sigma} N_{\nu}{}^{\rho \sigma}-\frac 14\, g_{\mu \nu}\,C_{ \delta \rho \sigma}N^{\delta \rho \sigma}\,.
\end{equation} 
The total energy-momentum conservation law takes the form
\begin{equation}\label{16}
\frac{\partial}{\partial x^\mu}\,\Big[\sqrt{-g}\,(T_{\hat{\alpha}}{}^\mu + \mathcal{T}_{\hat{\alpha}}{}^\mu -\frac{\Lambda}{\kappa}\,e^\mu{}_{\hat{\alpha}})\Big]=0\,.
 \end{equation}
 
 It is interesting to see how $N_{\mu \nu \rho}$ modifies GR field equations; to this end, we substitute
 \begin{equation}\label{17}
\mathfrak{H}_{\mu \nu \rho} = \mathcal{H}_{\mu \nu \rho} - \frac{\sqrt{-g}}{\kappa} N_{\mu \nu \rho}\,
\end{equation}
 in the Einstein tensor~\eqref{8} and employ modified TEGR field Equation~\eqref{13} to get 
 \begin{equation}\label{18}
^{0}G_{\mu \nu} + \Lambda g_{\mu \nu} = \kappa T_{\mu \nu}   +  Q_{\mu \nu} -  \mathcal{N}_{\mu \nu}\,,
\end{equation}
where $\mathcal{N}_{\mu \nu}$ is a  tensor defined by
\begin{equation}\label{19}
\mathcal{N}_{\mu \nu} := g_{\nu \alpha} e_\mu{}^{\hat{\gamma}} \frac{1}{\sqrt{-g}} \frac{\partial}{\partial x^\beta}\,(\sqrt{-g}N^{\alpha \beta}{}_{\hat{\gamma}})\,.
\end{equation} 
Therefore, to find the field equations of modified GR, we must add $Q_{\mu \nu} - \mathcal{N}_{\mu \nu}$ to the right-hand side of Einstein's field equations of GR.

Finally, we have to relate $N_{\mu \nu \rho} = - N_{\nu \mu \rho}$ to the torsion tensor. In NLG, the components of $N_{\mu \nu \rho}$ measured by the preferred observers of the theory with adapted tetrads $e^\mu{}_{\hat{\alpha}}$ are associated with the corresponding measured components of $X_{\mu \nu \rho}$ that are directly connected to the torsion tensor, and its expression has been discussed in detail in~\cite{BMB}. That is~\cite{Puetzfeld:2019wwo, Mashhoon:2022ynk}, 
\begin{equation}\label{20}
N_{\hat \mu \hat \nu \hat \rho}(x) = \int \mathcal{K}(x, x')\,X_{\hat \mu \hat \nu \hat \rho }(x') \sqrt{-g(x')}\, d^4x'\,,
\end{equation}
where
\begin{equation}\label{21}
X_{\hat \mu \hat \nu \hat \rho}= \mathfrak{C}_{\hat \mu \hat \nu \hat \rho}+ \check{p}\,(\check{C}_{\hat \mu}\, \eta_{\hat \nu \hat \rho}-\check{C}_{\hat \nu}\, \eta_{\hat \mu \hat \rho})\,.
\end{equation}
Here,  $\mathcal{K}(x, x')$ is the basic causal kernel of NLG that in essence must be determined via observation~\cite{BMB}, $\check{p}\ne 0$ is a constant dimensionless parameter, and 
 $\check{C}^\mu$ is the torsion pseudovector, 
\begin{equation}\label{22}
\check{C}_\mu =\frac{1}{3!} C^{\alpha \beta \gamma}\,E_{\alpha \beta \gamma \mu}\,,
\end{equation}
where $E_{\alpha \beta \gamma \delta}$ is the Levi-Civita tensor. 

Nonlocal gravity (NLG) is thus a classical extension of GR that is highly nonlinear as well. Linearized NLG has been investigated in detail~\cite{BMB}. Within the Newtonian regime of NLG, it appears possible to account for the rotation curves of nearby spiral galaxies as well as for the solar system data~\cite{Rahvar:2014yta, Chicone:2015coa, Roshan:2021ljs, Roshan:2022zov, Roshan:2022ypk}. Beyond the linear domain, no exact solution is known except for the trivial result that in the absence of gravity we have Minkowski spacetime~\cite{Bini:2016phe}. On the other hand, it is possible that certain nonlinear features of NLG that belong to the strong-field regimes such as those involving black holes or cosmological models may indeed survive in the local limit of the theory. It is therefore interesting to explore this limiting case of NLG. 

To come up with the local limit of NLG, let us assume that the kernel in Equation~\eqref{20} is proportional to the 4D Dirac delta function, namely,
\begin{equation}\label{23}
\mathcal{K}(x, x') := \frac{S(x)}{\sqrt{-g(x)}}\,\delta(x-x')\,;
\end{equation}
then, $N_{\mu \nu \rho}(x) = S(x) X_{\mu \nu \rho}$, where $S(x)$ is a dimensionless scalar function. Therefore,   
\begin{equation}\label{24}
N_{\mu \nu \rho}(x) = S(x)\,[\mathfrak{C}_{\mu \nu \rho}(x) + \check{p}\,(\check{C}_\mu\, g_{\nu \rho}-\check{C}_\nu\, g_{\mu \rho})]\,
\end{equation}
and the constitutive relation takes the form
\begin{equation}\label{25}
\mathcal{H}_{\mu \nu \rho} = \frac{\sqrt{-g}}{\kappa}[(1+S)\,\mathfrak{C}_{\mu \nu \rho}+ S\,\check{p}\,(\check{C}_\mu\, g_{\nu \rho}-\check{C}_\nu\, g_{\mu \rho})]\,.
\end{equation}
Here, the susceptibility function $S(x)$ is a characteristic of the background spacetime just as $\epsilon(x)$ and $\mu(x)$ are features of the medium in electrodynamics. In general,  the local electric permittivity $\epsilon(x)$ and magnetic permeability $\mu(x)$ functions are expected to preserve significant features of the electrodynamics of media such as spatial symmetries and temporal dependence. Similarly, $S(x)$ is expected to preserve the characteristics of the background spacetime.  Ultimately, $S(x)$ must be determined based on observational data~\cite{Tabatabaei:2022tbq, Tabatabaei:2023qxw}.

For $S(x) = 0$, we recover TEGR; otherwise, we have a natural generalization of GR that contains a new function $S(x)$.   Indeed, Equation~\eqref{25} implies that to have GR as a limit, we must impose the requirement that $1+S > 0$. In this local limit of nonlocal gravity, explicit deviations from locality have vanished; however, nontrivial aspects of NLG may have survived through $S(x)$ that would be interesting to study. Consequently, we explore the cosmological implications of this local limit of NLG. Spatially homogeneous and isotropic (FLRW) cosmological models have been treated in~\cite{Tabatabaei:2022tbq, Tabatabaei:2023qxw} in connection with $H_0$ tension. Therefore, we concentrate here on a class of spatially homogeneous but anisotropic spacetimes. 
 
 
\section{Anisotropic Models}

Let us consider a Bianchi type I model with a metric of the form
\begin{align}\label{K1}
ds^2=-dt^2+X^2 dx^2+Y^2dy^2+Z^2 dz^2, 
\end{align}
where $X$, $Y$ and $Z$ are  functions of time $ t$. This spacetime is spatially homogeneous with three spacelike commuting  Killing vector fields  $\partial_x$,   $\partial_y$, and  $\partial_z$.  A detailed discussion of such spacetimes is contained in Chapter 13 of~\cite{SKM}; for a recent discussion within the context of teleparallelism, see~\cite{Coley:2023ibm}.

Einstein's gravitational field equations  are
\begin{align}\label{K2}
^0G_{\mu \nu} +  \Lambda\, g_{\mu \nu} =8\pi G\, T_{\mu\nu}\,, \qquad ^0G_{\mu \nu} = ~^0R_{\mu\nu}-\frac{1}{2} g_{\mu\nu} \,^0R\,,
\end{align}
where  $T_{\mu\nu}$ is assumed to be due to the presence of a comoving perfect fluid with density $\rho(t)$ and pressure $P(t)$,
\begin{align}\label{K3}
T_{\mu\nu}=(\rho+P) U_\mu U_\nu+P g_{\mu\nu},
\end{align}
and $\Lambda$ is the cosmological constant.
With respect to the system of coordinates $x^\mu = (t, x, y, z)$, the perfect fluid is comoving with  $U^\mu   = \delta^\mu{}_{0}$; hence, 
\begin{align}\label{K4}
T_{\mu\nu} = {\rm diag}(\rho, P X^2, PY^2, PZ^2)\,.
\end{align}
Moreover, the Einstein tensor $^0G_{\mu \nu}$ is diagonal as well with components
\begin{align}\label{K5}
^0G_{00}& = \frac{\dot X\dot Y}{XY}+\frac{\dot Y\dot Z}{YZ}+\frac{\dot Z\dot X}{ZX}\,,\\ 
\label{K6} ^0G_{11}& = -X^2\left(\frac{\ddot Y}{Y}+\frac{\ddot Z}{Z}+\frac{\dot Y\dot Z}{YZ}\right)\,,\\
\label{K7} ^0G_{22}& = -Y^2\left(\frac{\ddot Z}{Z}+\frac{\ddot X}{X}+\frac{\dot Z\dot X}{ZX}\right)\,,\\
\label{K8} ^0G_{33}& = -Z^2\left( \frac{\ddot X}{X}+\frac{\ddot Y}{Y}+\frac{\dot X\dot Y}{XY}\right)\,.
\end{align}

It is interesting to work out the Kretschmann scalar $\mathcal{K}$,
\begin{align}\label{K8a}
\mathcal{K} = ~^0R_{\mu \nu \rho \sigma}\,^0R^{\mu \nu \rho \sigma}\,,
\end{align}
for metric~\eqref{K1}. The result is
\begin{align}\label{K8b}
\frac{1}{4} \,\mathcal{K} = \left( \frac{\ddot X}{X}\right)^2 + \left( \frac{\ddot Y}{Y}\right)^2 + \left( \frac{\ddot Z}{Z}\right)^2 + \left(\frac{\dot X\dot Y}{XY}\right)^2 +  \left(\frac{\dot Y\dot Z}{YZ}\right)^2 +  \left(\frac{\dot Z\dot X}{ZX}\right)^2\,.
\end{align}

Detailed discussions of the GR solutions of these models with $\Lambda = 0$ for dust ($P = 0$) can be found, for instance, in~\cite{HS1, HS2},  Section 5.4 of Ref.~\cite{HE}, and Section 12.15 of Ref.~\cite{PK}. We give a brief description of these solutions in Section IV in connection with cosmic deceleration. 

We are interested in the extended GR framework. Therefore, consider the class of observers that are spatially at rest  with adapted tetrad $e^{\mu}{}_{\hat \alpha}$ field given by 
\begin{align}\label{K9}
e^{\mu}{}_{\hat 0} = (1, 0, 0, 0)\,, \quad e^{\mu}{}_{\hat 1} = (0, \frac{1}{X}, 0, 0)\,, \quad e^{\mu}{}_{\hat 2} = (0, 0,\frac{1}{Y}, 0)\,, \quad e^{\mu}{}_{\hat 3} = (0, 0, 0, \frac{1}{Z})\,,
\end{align}
where the spatial axes point along the Cartesian coordinate directions. We have
\begin{align}\label{K10}
e_{\mu}{}^{\hat 0} = (1, 0, 0, 0)\,, \quad e_{\mu}{}^{\hat 1} = (0, X, 0, 0)\,, \quad e_{\mu}{}^{\hat 2} = (0, 0,Y, 0)\,, \quad e_{\mu}{}^{\hat 3} = (0, 0, 0, Z)\,.
\end{align}
We compute the Weitzenb\"ock torsion tensor~\eqref{4} in this case and we find $C_{\mu \nu}{}^{0} = 0$, $C_{i j}{}^{k} = 0$, and the only nonzero components can be obtained from 
\begin{align}\label{K11}
C_{0 1}{}^{1} = \frac{\dot{X}}{X}\,, \qquad C_{0 2}{}^{2} = \frac{\dot{Y}}{Y}\,, \qquad C_{0 3}{}^{3} = \frac{\dot{Z}}{Z}\,.
\end{align}
Similarly, we have $C_{\mu \nu 0} = 0$, $C_{i j k} = 0$, and the only nonzero components of $C_{\mu \nu \rho}$ can be obtained from 
\begin{align}\label{K12}
C_{0 11} = \dot{X}\,X\,, \qquad C_{0 22} = \dot{Y}\,Y\,, \qquad C_{0 33} = \dot{Z}\,Z\,.
\end{align}
It follows from these results that the torsion vector is given by
\begin{align}\label{K13}
C_{0} = -\left(\frac{\dot{X}}{X}+\frac{\dot{Y}}{Y}+\frac{\dot{Z}}{Z}\right)\,, \qquad C_i = 0\,, 
\end{align}
while the torsion pseudovector $\check{C}_\mu = 0$ in this case. 

The calculations of contorsion~\eqref{7} and the auxiliary torsion~\eqref{9} tensors produce similar results. That is,  $K_{0\mu \nu} = 0$, $K_{ijk} = 0$, and the only nonzero components of $K_{\mu \nu \rho}$ can be obtained from 
\begin{align}\label{K14}
K_{1 0 1} = \dot{X}\,X\,, \qquad K_{2 0 2} = \dot{Y}\,Y\,, \qquad K_{3 0 3} = \dot{Z}\,Z\,.
\end{align}
Moreover, $\mathfrak{C}_{\mu \nu 0} = 0$, $\mathfrak{C}_{ijk} = 0$, and the only nonzero components of $\mathfrak{C}_{\mu \nu \rho}$ can be obtained from
\begin{align}\label{K15}
\mathfrak{C}_{10 1} = X^2\,\left(\frac{\dot{Y}}{Y}+\frac{\dot{Z}}{Z}\right)\,, \quad \mathfrak{C}_{20 2} = Y^2\,\left(\frac{\dot{X}}{X}+\frac{\dot{Z}}{Z}\right)\,, \quad \mathfrak{C}_{3 03} =  Z^2\,\left(\frac{\dot{X}}{X}+\frac{\dot{Y}}{Y}\right)\,.
\end{align}

In the local limit of NLG, the constitutive relation of modified TEGR is given by $N_{\mu \nu \rho}(x) = S(x) \mathfrak{C}_{\mu \nu \rho}(x)$, where the gravitational susceptibility $S$ is a property of the background spacetime. In the case of the homogeneous time-dependent background~\eqref{K1}, we assume that $S$ is a function of time $t$. Therefore, $N_{\mu \nu 0} = 0$, $N_{ijk} = 0$ and the only nonzero components of $N_{\mu \nu \rho}$ can be obtained from
\begin{align}\label{K16}
N_{10 1} = S(t)\,X^2\,\left(\frac{\dot{Y}}{Y}+\frac{\dot{Z}}{Z}\right)\,, \quad N_{20 2} = S(t)\,Y^2\,\left(\frac{\dot{X}}{X}+\frac{\dot{Z}}{Z}\right)\,, \quad N_{3 03} =  S(t)\,Z^2\,\left(\frac{\dot{X}}{X}+\frac{\dot{Y}}{Y}\right)\,.
\end{align}

We can now compute $Q_{\mu \nu}$ given in Equation~\eqref{14} and $\mathcal{N}_{\mu \nu}$ given in Equation~\eqref{18}. The results are that these quantities are diagonal with elements
\begin{align}\label{K17}
Q_{00} = -S(t)\left(\frac{\dot X\dot Y}{XY}+\frac{\dot Y\dot Z}{YZ}+\frac{\dot Z\dot X}{ZX}\right)\,,
\end{align}
\begin{align}\label{K18}
Q_{11} = -S(t) X^2\,\frac{\dot Y\dot Z}{YZ}\,, \qquad Q_{22} = -S(t)Y^2\,\frac{\dot Z\dot X}{ZX}\,, \qquad Q_{33} = -S(t)Z^2\,\frac{\dot X\dot Y}{XY}\,.
\end{align}
For $\mathcal{N}_{\mu \nu}$, however, we find $\mathcal{N}_{00} = 0$ and 
\begin{align}\label{K19}
\mathcal{N}_{11} &= - \frac{X^2}{YZ}\, \frac{d}{dt}[S(Y\dot{Z} + \dot{Y}Z)]\,,\\
\label{K20}\mathcal{N}_{22}& = - \frac{Y^2}{XZ}\, \frac{d}{dt}[S(X\dot{Z} + \dot{X}Z)]\,, \\  
\label{K21} \mathcal{N}_{33} &= - \frac{Z^2}{XY}\, \frac{d}{dt}[S(X\dot{Y} + \dot{X}Y)]\,.
\end{align}

Collecting everything, the modified GR field Equation~\eqref{18} can be expressed as
\begin{align}\label{K22}
(1+S)\left(\frac{\dot X\dot Y}{XY}+\frac{\dot Y\dot Z}{YZ}+\frac{\dot Z\dot X}{ZX}\right) = \Lambda + 8 \pi G \rho\,
\end{align}
and 
\begin{align}\label{K23}
(1+S)\left(\frac{\ddot Y}{Y}+\frac{\ddot Z}{Z}+ \frac{\dot Y\dot Z}{YZ}\right) &= \Lambda - 8 \pi G P-\frac{dS}{dt}\,\left(\frac{\dot{Y}}{Y}+\frac{\dot{Z}}{Z}\right)\,,  \\
\label{K24}(1+S)\left(\frac{\ddot X}{X}+\frac{\ddot Z}{Z}+ \frac{\dot X\dot Z}{XZ}\right) &= \Lambda - 8 \pi G P-\frac{dS}{dt}\,\left(\frac{\dot{X}}{X}+\frac{\dot{Z}}{Z}\right)\,,  \\
\label{K25}(1+S)\left(\frac{\ddot X}{X}+\frac{\ddot Y}{Y}+ \frac{\dot X\dot Y}{XY}\right) &= \Lambda - 8 \pi G P-\frac{dS}{dt}\,\left(\frac{\dot{X}}{X}+\frac{\dot{Y}}{Y}\right)\,.
\end{align}

\subsection{Field Equations}

To express the gravitational field equations for the anisotropic models under consideration in a more tractable form, it is useful to consider
\begin{align}\label{K26}
V(t) = XYZ\,, \qquad W(t) := \frac{\dot X\dot Y}{XY}+\frac{\dot Y\dot Z}{YZ}+\frac{\dot Z\dot X}{ZX}\,,
\end{align}
where $|V(t)| = \sqrt{-g}$ and note that 
\begin{align}\label{K27}
\frac{\dot{V}}{V} = \frac{\dot X}{X} + \frac{\dot Y}{Y}+\frac{\dot Z}{Z}\,,\qquad \frac{\ddot V}{V} = \frac{\ddot X}{X} + \frac{\ddot Y}{Y}+\frac{\ddot Z}{Z} + 2 \,W\,.
\end{align}
Let us add Equations~\eqref{K23}--\eqref{K25} to get  
\begin{align}\label{K28}
(1+S) \left(2\,\frac{\ddot V}{V} -3 W\right) = 3(\Lambda  - 8 \pi G P) - 2 \frac{dS}{dt}\,\frac{\dot{V}}{V}\,.
\end{align}
Using Equation~\eqref{K22}, we can write 
\begin{align}\label{K29}
(1+S)\,\frac{\ddot V}{V} = 3[\Lambda  + 4 \pi G(\rho - P)] -  \frac{dS}{dt}\,\frac{\dot{V}}{V}\,
\end{align}
or
\begin{align}\label{K30}
\frac{d}{dt}[(1+S)\,\dot V] = 3 V\,[\Lambda  + 4 \pi G(\rho - P)]\,. 
\end{align}

Another interesting result is obtained by writing Equation~\eqref{K22} as $VW = (\Lambda + 8 \pi G \rho)V/(1+S)$ and taking the time derivative of both sides. From the relation
\begin{align}\label{K31}
\frac{1}{V}\,\frac{d (VW)}{dt} =  \frac{\dot X}{X}\,\left(\frac{\ddot Y}{Y}+\frac{\ddot Z}{Z}+ \frac{\dot Y\dot Z}{YZ}\right)+ \frac{\dot Y}{Y}\,\left(\frac{\ddot X}{X}+\frac{\ddot Z}{Z}+ \frac{\dot X\dot Z}{XZ}\right)+\frac{\dot Z}{Z}\,\left(\frac{\ddot X}{X}+\frac{\ddot Y}{Y}+ \frac{\dot X\dot Y}{XY}\right)\,,
\end{align}
we find
\begin{align}\label{K32}
(1+S) \frac{d (VW)}{dt} = (\Lambda - 8 \pi G P)\dot{V} -2 \, \frac{dS}{dt} VW\,.
\end{align}
Using Equation~\eqref{K22}, we finally get
\begin{align}\label{K33}
\frac{d\rho}{dt} = - (\rho + P)\frac{\dot V}{V}  - \frac{\frac{dS}{dt}}{(1+S)}\,\left(\rho +\frac{\Lambda}{8\pi G}\right)\,.
\end{align}
Equations~\eqref{K30} and~\eqref{K33} are important consequences of the modified field equations.  

Furthermore, let us define a new temporal variable $\tau$  by
\begin{align}\label{K34}
\tau := \int_0^t \frac{dt'}{1+S(t')}\,, \qquad  dt = (1+S) d\tau\,;
\end{align}
hence, the spacetime metric in $(\tau, x, y, z)$ coordinates is
\begin{align}\label{K35}
 ds^2 = - (1+S)^2 d \tau^2  + X^2(\tau) dx^2 + Y^2(\tau) dy^2  + Z^2(\tau) dz^2\,,
\end{align}
where $S$ is now considered, by an abuse of notation, a function of $\tau$. For instance, let us suppose $S(t)  = t$; then, $\tau = \ln (1+t)$ and in the above metric we have in this case $S = -1 +e^\tau$. For metric~\eqref{K35}, the Kretschmann scalar $\mathcal{K}$ given by Equation~\eqref{K8b} can be expressed in terms of the new temporal variable $\tau$ using  
 \begin{align}\label{K36}
\frac{\dot{X}}{X} = (1+S)^{-1} \frac{1}{X}\,\frac{dX}{d\tau}\,
\end{align}
and
\begin{align}\label{K37}
\frac{\ddot{X}}{X}  = (1+S)^{-2} \,  \left(\frac{1}{X}\,\frac{d^2 X}{d \tau^2}- \frac{\frac{dS}{d\tau}}{1+S}\, \frac{1}{X}\,\frac{dX}{d\tau}\right)\,.
\end{align}
The gravitational field equations can now be written in terms of the temporal variable $\tau$ as
\begin{align}\label{K38}
\frac{1}{XY}\,\frac{dX}{d\tau}\,\frac{dY}{d\tau} + \frac{1}{YZ}\,\frac{dY}{d\tau}\,\frac{dZ}{d\tau} + \frac{1}{ZX}\,\frac{dZ}{d\tau}\,\frac{dX}{d\tau}= (1+S)(\Lambda + 8 \pi G \rho)\,,
\end{align}
\begin{align}\label{K39}
\frac{1}{Y}\,\frac{d^2 Y}{d \tau^2} + \frac{1}{Z}\,\frac{d^2 Z}{d \tau^2} + \frac{1}{YZ}\,\frac{dY}{d\tau}\,\frac{dZ}{d\tau} = (1+S)(\Lambda - 8 \pi G P)\,,
\end{align}
\begin{align}\label{K40}
\frac{1}{X}\,\frac{d^2 X}{d \tau^2} + \frac{1}{Z}\,\frac{d^2 Z}{d \tau^2} + \frac{1}{XZ}\,\frac{dX}{d\tau}\,\frac{dZ}{d\tau} = (1+S)(\Lambda - 8 \pi G P)\,,
\end{align}
\begin{align}\label{K41}
\frac{1}{X}\,\frac{d^2 X}{d \tau^2} + \frac{1}{Y}\,\frac{d^2 Y}{d \tau^2} + \frac{1}{XY}\,\frac{dX}{d\tau}\,\frac{dY}{d\tau} = (1+S)(\Lambda - 8 \pi G P)\,.
\end{align}

To solve Equations~\eqref{K39}--\eqref{K41}, let us subtract, for instance, Equation~\eqref{K39} from Equation~\eqref{K40} to get 
\begin{align}\label{K42}
\frac{1}{X}\,\frac{d^2 X}{d \tau^2} - \frac{1}{Y}\,\frac{d^2 Y}{d \tau^2} + \frac{1}{Z}\,\frac{dZ}{d \tau}\left(\frac{1}{X}\,\frac{dX}{d\tau} - \frac{1}{Y}\,\frac{dY}{d\tau}\right) = 0\,,
\end{align}
which with $V = XYZ$ can be written as
\begin{align}\label{K43}
\frac{d}{d\tau} \left(\frac{1}{X}\,\frac{dX}{d\tau} - \frac{1}{Y}\,\frac{dY}{d\tau}\right) + \frac{1}{V}\,\frac{dV}{d \tau}\,\left(\frac{1}{X}\,\frac{dX}{d\tau} - \frac{1}{Y}\,\frac{dY}{d\tau}\right) = 0\,.
\end{align}
Hence, 
\begin{align}\label{K44}
V\, \left(\frac{1}{X}\,\frac{dX}{d\tau} - \frac{1}{Y}\,\frac{dY}{d\tau}\right) = \mathcal{C}_{12}\,,
\end{align}
where $ \mathcal{C}_{12}$ is a constant of integration and similar results hold for the other metric functions. 

Finally, in terms of temporal variable $\tau$, Equation~\eqref{K30} can be written as
\begin{align}\label{K45}
 \frac{1}{V}\, \frac{d^2V}{d\tau^2} = 3(1+S)[\Lambda  + 4 \pi G(\rho - P)]\,.
\end{align}


\subsection{Special Solutions}

We now explore some cases of particular interest.

\subsubsection{de Sitter}

Let us first consider de Sitter's solution with
\begin{align}\label{S1}
\rho = P = 0\,, \qquad X = Y =Z = e^{\lambda t}\,, 
\end{align}
where $\lambda$ is a nonzero constant. The field equations imply
\begin{align}\label{S2}
3 \lambda^2 (1+S) = \Lambda\,, \qquad 3 \lambda^2 (1+S) = \Lambda - 2\lambda \frac{dS}{dt}\,. 
\end{align}
Therefore, $dS/dt = 0$ and $S$ must be constant. It follows that de Sitter spacetime is not a solution of the modified TEGR since the susceptibility function is independent of time while the background spacetime is dynamic. This is in agreement with the fact that de Sitter spacetime is not a solution of NLG~\cite{Mashhoon:2022ynk}. 

\subsubsection{Kasner}

Let us next consider the standard Kasner metric~\cite{Kas, Chicone:2011ie}
\begin{equation}\label{S3}
ds^2=-dt^2+ t^{2p_1}dx^2+ t^{2 p_2} dy^2+ t^{2 p_3} dz^2\,,
\end{equation}
\begin{equation}\label{S4}
p_1+p_2+p_3=p_1^2+p_2^2+p_3^2=1\,.
\end{equation}
In GR, this empty universe model is a solution of the field equations with $\Lambda = 0$ and $\rho = P = 0$.  Note that with $p_1 = 1$, say, and hence $p_2=p_3 = 0$, we recover flat spacetime. Therefore, we can assume     $p_1<p_2<p_3$;                    that is,
\begin{equation}\label{S5}
-\frac{1}{3}\le p_1\le 0,\qquad 0\le p_2\le \frac{2}{3},\qquad \frac{2}{3}\le p_3\le 1\,.
\end{equation}

It follows from field equations~\eqref{K22} and~\eqref{K23} that
\begin{equation}\label{S6}
\frac{(1+S)}{t^2} (p_1p_2 +p_2p_3+p_1p_3) = \Lambda + 8 \pi G \rho\,, 
\end{equation}
\begin{equation}\label{S7}
\frac{(1+S)}{t^2} (p_2^2 - p_2 + p_3^2 - p_3 + p_2p_3) = \Lambda - 8 \pi G P -\frac{dS}{dt}(p_2 + p_3)\,, 
\end{equation}
etc. Because $p_1p_2 +p_2p_3+p_1p_3 = 0$ and $p_2^2 - p_2 + p_3^2 - p_3 + p_2p_3 = 0$ together with two other cyclically related terms that vanish, we find that $\rho$, $P$, and $S$ must be constants such that  
\begin{equation}\label{S8}
- 8 \pi G \rho = \Lambda \le 0\,, \qquad \rho + P = 0\,, \qquad  S = {\rm constant}\,.
\end{equation}
Therefore, Kasner's spacetime with constant $S$ is not a solution of the modified TEGR.  

\subsubsection{Flat FLRW Model}

We now consider $X = Y = Z = a(t)$. Then, field equations~\eqref{K22} and~\eqref{K23} imply
\begin{equation}\label{S9}
3(1+S)\left(\frac{\dot{a}}{a}\right)^2 = \Lambda + 8 \pi G \rho\,, 
\end{equation}
\begin{equation}\label{S10}
(1+S)\left[ 2\frac{\ddot{a}}{a} + \left(\frac{\dot{a}}{a}\right)^2\right] = \Lambda - 8 \pi G P - 2\frac{dS}{dt}\,\frac{\dot{a}}{a}\,. 
\end{equation}
A thorough treatment of these equations is contained in a recent paper~\cite{Tabatabaei:2022tbq}, where they were employed with $\Lambda = 0$ in a detailed discussion of the implications of the modified Cartesian flat cosmology in connection with $H_0$ tension. 

It is important to emphasize that in these time-dependent solutions considered thus far, $S$ must be dependent upon time as well; otherwise, we do not have a physically meaningful solution of the theory. 


\section{Solution for Dust with $\Lambda = 0$}


\subsection{$S = 0$}

When $S = 0$, we get the gravitational field equations in GR. A well-known class of anisotropic solutions corresponds to $\Lambda = 0$ and $P = 0$~\cite{HS1, HS2, HE, PK}. In this case, Equations~\eqref{K30} and~\eqref{K33} imply 
\begin{align}\label{D1}
\rho \,V = \frac{\ell_0}{ 6 \pi G}\,, \qquad V = \ell_0\, t^2(1 + \Sigma/t)\,, 
\end{align}
where $\ell_0 > 0$ and $\Sigma > 0$ are integration constants with dimensions of length (or time, since $c = 1$) and we have assumed that $V = 0$ at $t = 0$. Using
\begin{align}\label{D2}
\int \frac{dt}{V} = -\frac{1}{\ell_0 \Sigma} \ln(1+ \Sigma /t)\,, 
\end{align}
Equation~\eqref{K44} and similar ones can be integrated with the result that ratios such as $X/Y$, etc., up to constant coefficients are given by $(1+ \Sigma /t)$ to some constant powers. In the absence of anisotropy ($\Sigma = 0$), we must recover the standard Einstein-de Sitter solution; that is, $\Sigma$ is the anisotropy parameter for finite $t > 0$.  Therefore, we look for the solutions of the field equations such that the metric coefficients are given by
\begin{align}\label{D3}
X = \ell_0{}^{q_1}\,t^{2/3}(1 + \Sigma/t)^{q_1}\,, \quad  Y = \ell_0{}^{q_2}\,t^{2/3}(1 + \Sigma/t)^{q_2}\,, \quad  Z = \ell_0{}^{q_3}\,t^{2/3}(1 + \Sigma/t)^{q_3}\,, 
\end{align}
where $q_i$, $i = 1, 2, 3$, are constants that must add up to unity in order to satisfy Equation~\eqref{D1}. A detailed examination reveals that the field equations are all satisfied in the case of dust with $\Lambda = 0$ provided
\begin{align}\label{D4}
q_1 + q_2 + q_3 = 1\,, \qquad  q_1^2 + q_2^2 + q_3^2 = 1\,. 
\end{align}

It is possible to choose $q_1<q_2<q_3$; that is, $-\frac{1}{3}\le q_1\le 0$, $0\le q_2\le \frac{2}{3}$, and $\frac{2}{3}\le q_3\le 1$. Then,  a convenient representation of these constants  is given by
\begin{align}\label{D5}
q_i = \frac{1}{3} -\frac{2}{3} \sin [\theta - 2(i-1) \pi/3]\,, \qquad i = 1, 2, 3\,, \qquad  \frac{\pi}{6} \le \theta \le \frac{\pi}{2}\,.
\end{align}

It is important to note that at late times, $t \to \infty$, we recover the standard Einstein-de Sitter solution.  


\subsection{$S\ne 0$}

Let us note that a constant $S \ne 0$ is equivalent to a constant rescaling of the time coordinate in GR, namely, $t \to (1+S)^{1/2} t$. Therefore, new results are obtained only when $S$ depends upon time, which is necessary for a proper physical interpretation of the susceptibility function $S$. 

We now consider the case $dS/dt \ne 0$.  For $P = 0$ and $\Lambda = 0$, it follows from Equations~\eqref{K33} and~\eqref{K45} that  
\begin{align}\label{D6}
(1+S) \rho \,V = \frac{\ell_0}{ 6 \pi G}\,, \qquad V = \ell_0\, \tau^2(1 + \Sigma/\tau)\,. 
\end{align}
Moreover, the metric coefficients are given by
\begin{align}\label{D7}
X = \ell_0{}^{q_1}\,\tau^{2/3}(1 + \Sigma/\tau)^{q_1}\,, \quad  Y = \ell_0{}^{q_2}\,\tau^{2/3}(1 + \Sigma/\tau)^{q_2}\,, \quad  Z = \ell_0{}^{q_3}\,\tau^{2/3}(1 + \Sigma/\tau)^{q_3}\,, 
\end{align}
where $q_1 + q_2 + q_3 = 1$ and  $q_1^2 + q_2^2 + q_3^2 = 1$. 
For $\Sigma = 0$ or $\tau \to \infty$, we recover the isotropic solution corresponding to the modified Einstein-de Sitter model. 

It is interesting to explore the implications  of these solutions for the dimensionless deceleration parameter defined for the $x$ direction, say, by $Q_x = - (\ddot{X} / X)/ (\dot{X}/X)^2$, etc.  

Let us first recall that $\Sigma > 0$ is the anisotropy parameter; in fact, for $\Sigma = 0$, we get the standard Einstein-de Sitter result that $Q = 1/2$ in every direction. Let anisotropic expansion occur in the $z$ direction; then, Equation~\eqref{D4} allows two possibilities, namely, $(q_1, q_2, q_3) = (0, 0, 1)$ and $(q_1, q_2, q_3) = (2/3, 2/3, -1/3)$.  We choose the former case for the sake of simplicity. We are interested in the nature of deceleration parameter $Q_z$ in the $z$ direction. For $S= 0$,  we have $Q_x = Q_y = 1/2$ and
\begin{align}\label{D8}
Q_z = 2 \frac{(t + \Sigma)(t-2\Sigma)}{(2t -\Sigma )^2} < \frac{1}{2}\,. 
\end{align}
Indeed, for $t < 2\,\Sigma$ we have $Q_z < 0$ and hence acceleration in the $z$ direction that later turns into deceleration for $t > 2\,\Sigma$ with magnitude $<1/2$. As $ t \to \infty$, $Q_z \to 1/2$; that is, isotropy is recovered at late times.

How do these results change in the presence of $S(t)$? In terms of temporal parameter $\tau$, we find for the $x$ direction, say, 
\begin{align}\label{D9}
Q_x  =  - X \frac{d^2X}{d\tau^2}\left(\frac{dX}{d\tau}\right)^{-2} + \frac{dS}{dt} X\left(\frac{dX}{d\tau}\right)^{-1}\,. 
\end{align}
Therefore, with 
\begin{align}\label{D10}
X = Y = \tau^{2/3}\,, \qquad  Z = \ell_0\,\tau^{2/3}(1 + \Sigma/\tau)\,,
\end{align}
the results are
\begin{align}\label{D11}
Q_x = Q_y =  \frac{1}{2} + \frac{3}{2} \tau\, \frac{dS}{dt}\, 
\end{align}
and
\begin{align}\label{D12}
Q_z = 2 \frac{(\tau + \Sigma)(\tau -2\Sigma)}{(2\tau  -\Sigma )^2} + 3 \left(\frac{\tau + \Sigma}{2\tau - \Sigma}\right)\tau\, \frac{dS}{dt}\,. 
\end{align}

Let us suppose, for instance, that $dS/dt > 0$. Then, the deceleration increases in the $x$ and $y$ directions. The same is true in the $z$ direction for $\tau >  2\,\Sigma$; however, for $\tau < \Sigma/2$, the presence of $dS/dt > 0$ causes extra acceleration in the $z$ direction. For $2\,\Sigma > \tau > \Sigma/2$, the sign of $Q_z$ depends upon the magnitude of $dS/dt$. We note that isotropy is restored at late times  $\tau \gg \Sigma$.

 
\section{Solution for Dynamic Dark Energy with $\Lambda = 0$}

Let us imagine a universe that in the absence of the cosmological constant ($\Lambda = 0$) is dominated by dynamic dark energy with $P_{de} + \rho_{de} = 0$. Therefore, Equation~\eqref{K33} implies 
\begin{align}\label{E1}
\rho_{de} = \rho_{de}(t_0) \,\frac{1+S_0}{1+S}\,, \qquad \rho_{de} (t_0) > 0\,, \quad S_0 = S(t_0)\,,
\end{align} 
where $t_0$ refers to some fiducial epoch in the expansion of the universe. The pressure of dark energy is always negative, $P_{de} = - \rho_{de}$, and we assume on the basis of Equation~\eqref{E1} that 
\begin{align}\label{E2}
6 \pi G \rho_{de} (1+S) =  6 \pi G \rho_{de}(t_0) \,(1+S_0) := \frac{1}{\tau_0^2}\,, \qquad  \eta := \frac{\tau}{\tau_0}\,,
\end{align} 
where $\tau_0 > 0$ is a constant with the dimensions of time and henceforth we employ $\eta$ as the new dimensionless temporal variable. To explore the anisotropic acceleration of this dark energy universe model, we assume for the sake of simplicity that $X = Y$. Thus we have a cylindrical model with the $z$ direction as the main direction of anisotropy. Let us note here that for $S = 0$ our dynamic dark energy source in effect reduces to a cosmological constant. 

With $X = Y$, Equation~\eqref{K41} reduces to 
\begin{align}\label{E3}
\frac{2}{X}\,\frac{d^2 X}{d \eta^2} + \frac{1}{X^2}\,\left(\frac{dX}{d\eta}\right)^2 = \frac{4}{3}\,.
\end{align}
This nonlinear equation can be easily solved with 
\begin{align}\label{E4}
X := \alpha^{2/3}\,, \qquad \frac{d^2 \alpha}{d\eta^2} - \alpha = 0\,,\qquad  \alpha = C_{+}\, e^\eta + C_{-}\, e^{-\eta}\,,
\end{align}
where $C_{\pm}$ are integration constants. Next, Equation~\eqref{K38} implies
\begin{align}\label{E5}
\frac{1}{X^2}\,\left(\frac{dX}{d\eta}\right)^2+ \frac{2}{Z}\,\frac{dZ}{d\eta} \frac{1}{X}\frac{dX}{d\eta}=\frac{4}{3}\,,
\end{align}
or
\begin{align}\label{E6}
\frac{1}{Z}\,\frac{dZ}{d\eta} = \frac{\alpha}{\beta} - \frac{1}{3} \frac{\beta}{\alpha}\,,
\end{align}
where $\alpha = d\beta/d\eta$ and 
\begin{align}\label{E7}
 \beta = \frac{d\alpha}{d\eta} = C_{+}\, e^\eta - C_{-}\, e^{-\eta}\,.
\end{align}
Hence, $Z = C_0\, \alpha^{-1/3}\, \beta$, where $C_0$ is an integration constant. Therefore, 
\begin{align}\label{E8}
X = Y =  \alpha^{2/3}\,,\quad  Z = C_0\, \alpha^{-1/3}\, \beta\,, \quad V = X^2 Z = C_0\, \alpha\,\beta  = C_0 ( C^2_{+}\, e^{2\eta} - C^2_{-}\, e^{-2\eta})\,.
\end{align}
One can check that this is the general solution of the field equations in the present case.  For these solutions, the Kretschmann scalar $\mathcal{K}$ can be written as 
\begin{align}\label{E8a}
\frac{1}{4} (1+S)^4 \,\tau_0^4\,\mathcal{K} = \frac{16}{9} K_1 -\frac{8}{3} \frac{\beta}{\alpha}  \left(1-\frac{\beta^2}{3\alpha^2}\right)\tau_0\,\frac{dS}{dt} + K_2\, \frac{\alpha^2}{\beta^2} \left(\tau_0\,\frac{dS}{dt}\right)^2\,,
\end{align}
where 
\begin{align}\label{E8b}
 K_1  = 1- \frac{2}{3}\frac{\beta^2}{\alpha^2} +\frac{1}{3} \frac{\beta^4}{\alpha^4}\,, \qquad  K_2  = 1- \frac{2}{3}\frac{\beta^2}{\alpha^2} + \frac{\beta^4}{\alpha^4}\,.
\end{align}
We are interested in the behavior of the acceleration of these anisotropic solutions.

For the deceleration parameters of this model, we find
\begin{align}\label{E9}
Q_x = Q_y = \frac{1}{2}\left(1-3\frac{\alpha^2}{\beta^2}\right) + \frac{3}{2}\tau_0 \frac{dS}{dt} \frac{\alpha}{\beta}\,, 
\end{align}
\begin{align}\label{E10}
 Q_z = -4\left(1-3\frac{\alpha^2}{\beta^2}\right)^{-2} - 3\tau_0 \frac{dS}{dt} \frac{\alpha}{\beta}\left(1-3\frac{\alpha^2}{\beta^2}\right)^{-1}\,.
\end{align}
At late times $\eta \to \infty$,  $\alpha/\beta \to 1$, and these expressions reduce to $Q_x = Q_y = Q_z = -1 + (3\tau_0/2)\, dS/dt$. However, for finite $\eta$ there could be anomalous behavior when $\beta = 0$
 or $3\alpha^2 = \beta^2$. If $\beta = 0$, then $Q_x = Q_y = -\infty$, while $Q_z = 0$. On the other hand, if $3\alpha^2 = \beta^2$, then $Q_x = Q_y = \pm (\sqrt{3} \tau_0/2) dS/dt$, while $Q_z = -\infty$.  To give an example of the former situation, consider, for instance, the case where $C_{+} = C_{-} = C > 0$ with $\alpha = 2C\, \cosh \eta$ and $\beta = 2C\, \sinh \eta$. In this case, $3\alpha^2 > \beta^2$. Then, $X = Y = (2C)^{2/3} \cosh^{2/3}\eta$ and $Z = C_0(2C)^{2/3} \cosh^{-1/3}\eta\,\sinh \eta$. The universe starts from a singular pancake state at $\eta = 0$ and expands with infinite acceleration in the $x$ and $y$ directions but with zero acceleration in the $z$ direction. Indeed, inspection of the Kretschmann scalar $\mathcal{K}$ given by Equation~\eqref{E8a} reveals that $\mathcal{K}$ diverges at $\tau = 0$ provided $dS/d\tau \ne 0$. 
 
 For an example of the case where $3\alpha^2 = \beta^2$, let us consider, for instance, $C_{+} = 1$ and $C_{-} = - (2+\sqrt{3})$. Then, the universe starts from a cigar state at $\eta = 0$ with infinite acceleration in the $z$ direction but with finite acceleration $-(\sqrt{3} \tau_0/2) dS/dt$ in the $x$ and $y$ directions assuming that $dS/dt > 0$.  The Kretschmann scalar~\eqref{E8a} remains finite in this case. 

These results could be interesting in connection with recent observational evidence in favor of anomalous  anisotropic acceleration of bulk flow in the local universe~\cite{Colin:2019opb, Secrest:2020has, Secrest:2022uvx, Solanki:2023yoa, Perivolaropoulos:2023tdt}. 

\section{DISCUSSION}

 The cosmic acceleration in the standard cosmological model is assumed to be isotropic. Large-scale anisotropy in the Hubble flow has been the subject of numerous studies.  We have explored  anisotropy in cosmic acceleration within the theoretical framework of a recent teleparallel extension of general relativity that corresponds to the local limit of nonlocal gravity. This modified theory involves a function $S(x)$ such that GR is recovered for $S = 0$. 
We are interested in the cosmological significance of the extra function $S(x)$. In particular, in a dynamic Bianchi type I model that is consistent with the modified GR field equations, we theoretically investigate  anisotropy in cosmic acceleration and determine the contribution of $S(t)$ to local anisotropic acceleration. Our results in Sections IV and V for possible local anisotropic cosmic acceleration depend explicitly upon $dS(t)/dt$. 

The function $S(t)$ may also possibly contribute to the resolution of an anomaly in the quadrupole anisotropy of CMB temperature. It has been reported that the CMB temperature angular power spectrum suffers from a deficit in its quadrupole moment~\cite{WMAP:2012fli,wmap,Planck:2019evm,Planck:2018vyg}. The anomaly has to do with the low amplitude of the quadrupole anisotropy compared to the prediction of the standard $\Lambda$CDM model. On the other hand, in an anisotropic universe described by metric~\eqref{K1}, one would in general expect a change in the quadrupole moment of the CMB temperature due to different amounts of redshift suffered by photons traveling after recombination along different directions toward the observer~\cite{Akarsu:2019pwn, Campanelli:2006vb, Campanelli:2007qn, Buniy:2005qm, Sachs:1967er}. However, Big Bang nucleosynthesis (BBN) puts a tight constraint on the anisotropy of the expansion rate~\cite{Barrow:1976rda}; hence, it is rather difficult for the anisotropic expansion to justify the CMB quadrupole deficit~\cite{Akarsu:2019pwn}. A suitable susceptibility function $S(t)$ of the modified gravity theory may be able to ameliorate the situation. That is, $S(t)$ may allow the  quadrupole deficit anomaly be alleviated while respecting the BBN constraint.  A more complete discussion is beyond the scope of the present paper.

\section*{ACKNOWLEDGMENTS}

The work of AB has been supported financially by Iran Science Elites Federation.

\appendix

\end{document}